\documentclass[12pt,a4paper]{article}
\usepackage{amssymb,amsmath,graphicx,subcaption,hyperref,cite,float,xcolor,soul,epstopdf}
\graphicspath{{Allee_effect_in_prey/}}
\usepackage{comment}
\usepackage{xcolor}
\usepackage{soul}
\usepackage{epstopdf}
\usepackage[utf8]{inputenc}
\usepackage[english]{babel}
\begin{document}
\begin{center}
   {\Large\bf A Study on the Specialist Predator with the Allee Effect on the Prey}
\end{center}
\begin{center} 
    Tanmay Das$^{a,*}$ and Mahatsab Mandal$^{b}$
   
   \textit{Department of Physics,Government General Degree College at Kalna-I}\\
   \textit{Muragacha, Medgachi, Purba Bardhdaman, Pin 713405,West Bengal, India}
   
   {Email$^a$:tanmay.physics@gmail.com}\\
   {Email$^{b}$:mahatsab@gmail.com}
\end{center}

\begin{abstract} 
Predator-prey models in theoretical ecology have 
a long and complex history, spanning decades of 
research. Most of the models rely upon simple 
reproduction and mortality rates associated with 
different types of functional responses. 
A key development in this field occurred with the 
introduction of a density dependent reproduction 
rate, originally introduced by Allee. 
In this manuscript, a new function representing 
the Allee effect is introduced and justified 
from the ecological point of view. 
This paper aims to analyze predator-prey models 
incorporating Holling type-I and II functional 
responses, influenced by this new Allee function. 
A rich dynamics shows up in the presence of the 
said function, including the emergence of the limit cycles through the Hopf bifurcation for a 
particular parameter domain.   
\end{abstract}



\section{Introduction}
Predator-prey models in mathematical biology have 
always provided interesting as well as complex 
outcomes starting from simpler interrelations 
between the species under consideration~\cite{May1973,  Crawley1992, Mueller2000, Neubert2002, Berryman1992, Kuang1998, selkov,strogatz,Das,Pita}.
It dates back to the Lotka-Volterra 
model~\cite{Lotka, Volterra} and incorporates 
many different functional responses and interspecies competition terms in the recent past. 
One key observation came from an ecologist W. C. Allee~\cite{Allee} 
in 1931, which proposed that the 
reproduction rate depends upon the per capita 
growth rate of the species. 
It may be caused by several factors, including 
difficulties in mate finding~\cite{Dennis1989}, 
social dysfunction, inbreeding 
depression~\cite{Courchamp2008}, 
predation~\cite{Gascoigne2004}, and food 
exploitation~\cite{Courchamp2006, Stephens1999}.
The Allee effect on prey populations usually destabilize predator-prey dynamics by increasing the risk of prey extinction~\cite{Sinclair1998, Courchamp1999, Courchamp2000, Kent2003, Zhou2005} and reducing the elasticity of the system. This destabilization can manifest in various ways. Firstly, a strong Allee effect can weaken predator-prey systems by disrupting equilibrium stability or slowing convergence to a stable state~\cite{Zhou2005, Wang2010}. This effect establishes a population threshold that must be surpassed for growth. On the other hand, weak Allee effects increase the risk of extinction to some extent, but the population can still survive at low densities~\cite{Courchamp2008, Boukal2007,Shi2006, Wang2001}. In recent years, many authors have studied the dynamical properties of the predator-prey model~\cite{Celik2008, Celik2009, Merdan2009, Gonzalez2010, Gonzalez2019, Flores2014, Sasmal2018, Sen2019, Sen2021, Ghosh2017, Kumar2022, Han2003} in the presence of the Allee effect.
In literature\cite{Merdan, Berec2007} pertinent to the present discussion, a general form of the Allee effect, depending upon the biological constraints, has been constructed, and a possible function representing the same has been proposed. 
This manuscript proposes a new function justifying 
the article by Merdan \cite{Merdan} and studies 
its impact on specialist predator-prey dynamics 
for two different functional responses, namely 
Holling type-I and II \cite{Holling1959, Holling1965}. 
In Holling Type-I, the linear increase suggests
that a predator's food processing time is 
negligible and that consuming prey does not hamper 
its search for additional food~\cite{Seo2008}. 
The Type-II functional response, also known as Holling’s disk equation, describes a scenario where the consumption rate decelerates as the prey population increases due to the predator's limited capacity for searching and processing prey. 
It is widely used in ecological modeling to investigate complex dynamics in predator-prey systems, including stable limit cycles and chaotic behavior~\cite{Gascoigne2004, Rosenzweig1963, Zhang2006, Zu2010, Zu2013, McLellan2010, Halder2019}.

The article is organized as follows. After a brief introduction to the Allee Effect, a new possible function representing the aforementioned phenomenon has been proposed in Section 2. Possible dynamical aspects of the newly proposed Allee affected prey population in the presence of Holling type-I coexistence with a specialist predator have been portrayed in Section 3. In Section 4, similar observations are drawn for Holling type-II functional response. The paper concludes with final remarks in Section 5.

\section{The Allee Effect}
Assuming $N$ and $P$ stand for the population of prey and specialist predator species, respectively it is possible to write a general form of the predator-prey dynamical system as:
\begin{subequations}
\begin{align}
\label{genmod1}
\frac{dN}{dT} &= f_1(N) N - g(N,P)\\
\frac{dP}{dT} &= \bar{\epsilon}\ g(N,P) -f_2(P) P
\end{align}
\end{subequations}
where, $f_1(N)$ and $f_2(P)$ stands for the per capita growth rate of prey and specialist predator, respectively, and $g(N,P)$ is the functional response depicting the food chain under consideration while $\bar{\epsilon}$ is the food conversion factor. In the presence of the Allee effect in prey population, the per capita growth rate may be split into two factors:
\begin{equation}
f_1(N) = \bar{\alpha}(N) \bar{f}_1(N)
\end{equation}    
where, $\bar{\alpha}(N)$ encodes the Allee effect and $\bar{f}_1(N)$ is the Allee free per capita growth rate of the prey population.  
In this manuscript, the logistic growth of prey has been considered with a carrying capacity $K$ and an intrinsic growth rate $\bar{r}$ to ensure the prey species sustains itself.
A few general properties of $\bar{\alpha}(N)$ may be chalked based on the ecological observations as follows:\\
\begin{gather}
\begin{split}
&(i)~~\bar{\alpha}(N) \ge 0\\
&(ii)~~\bar{\alpha}'(N) = \frac{d\bar{\alpha}(N)}{dN} \ge 0\\
&(iii)~~\lim\limits_{N \rightarrow \infty} \bar{\alpha}(N) = 1\\
&(iv)~~\bar{\alpha}(N = 0) = 0
\end{split}
\end{gather}
while, the first condition is self explanatory, the rest requires some enlightenment. $\bar{\alpha}'(N) \ge 0$ encodes the fact that the footprint of the Allee effect dilutes with increment in population. Additionally, for sufficiently large populations, finding suitable mates is much easier and effectively vanishes the Allee Effect. At the same time, for the opposite extreme, for sparsely populated species, the effect is stronger and turns the reproduction rate to its lowest possible value. These two constraints are reflected in the third and fourth conditions, respectively. In this manuscripts, a new function combining all the criteria for Allee effect has been proposed as: 
\begin{gather}
\bar{\alpha}(N) = 1 - \exp(-\bar{a} N)
\end{gather}
A tunable Allee parameter, $\bar{a}$, is introduced to allow for the adjustment of the Allee Effect. The functional response of a predator, $g(N,P)$, often takes the form  $AP\bar{g}(N)$, where $A$ represents the predator's attack rate. When $\bar{g}(N)=N$, it denotes a Holling type-I functional response or linear functional response, while $\bar{g}(N)=\frac{N}{1+\bar{h}N}$ stands for the Holling type-II functional response. 
Dynamical response due to the incorporation of above mentioned Allee function in the context of Holling type-I and II functional responses for two species interaction has been considered in the following sections of this manuscript.

\section{The Allee Effect in presence of Holling type-I functional response}
In the first subsection, the concerned model and its non-dimensionalization have been derived and then in subsequent subsections, dynamical aspects of the same will be studied.

The model under consideration is:
\begin{subequations}
	\begin{align}
	\frac{dN}{dT} &= \bar{r} N\bigg(1 - e^{-\bar{a} N}\bigg)\bigg(1 - \frac{N}{K}\bigg) - ANP\\
	\frac{dP}{dT} &= \bar{\epsilon} ANP - DP
	\end{align}
\end{subequations}
where, $f_2(P)=D$ is the intrinsic death rate of the predator, assumed to be constant. To non-dimensionalize this model, one can rescale the variables as:
\begin{gather}
x = \mu N, y = \nu P, t = \beta T
\end{gather}
After few simplifications, this substitution leads to 
\begin{subequations}
	\begin{align}
	\frac{dx}{dt} &=  \frac{\bar{r}}{\beta} x\bigg(1 - e^{-\frac{\bar{a} x}{\mu}} \bigg)\bigg(1 - \frac{x}{\mu K}\bigg) - \frac{A}{\beta \nu} xy\\
	\frac{dy}{dt} &= \bar{\epsilon}\frac{A}{\beta \mu}\,xy - \frac{D}{\beta}\,y 
	\end{align}
\end{subequations}
The normalization parameters then chosen to be 
\begin{gather*}
\mu = \frac{1}{K}, \beta = \bar{r}, \nu = \frac{A}{\bar{r}}
\end{gather*}
which leads to the non-dimensional model
\begin{subequations}
	\begin{align}
	\frac{dx}{dt} &= x\big(1 - e^{-ax}\big)\big(1 - x\big) - xy\\
	\frac{dy}{dt} &= \epsilon xy - d y
	\end{align}
\end{subequations}
where, few parameters have been re-labeled as $ a = \frac{\bar{a}}{\mu},\epsilon = \frac{\bar{\epsilon} AK}{\bar{r}}, d = \frac{D}{\bar{r}}$. Thus, the Allee function in non-dimensional form turns out to be 
\begin{gather}
\alpha(x) = 1 - e^{-ax}
\end{gather}
subject to the same constraints as imposed on 
$\bar{\alpha}(N)$.

\subsection{Local stability Analysis of the Model}
To analyze the system, it is ﬁrst required to locate ﬁxed points and then linearization about them
will reveal its stability around them. The ﬁxed points appearing at the boundary of the ﬁrst quadrant are distinguished from those located inside the same. For the co-existence of the species under consideration, it is the latter type that requires stability or must possess a limit cycle about them.

\subsubsection{Boundary Fixed points}
The system holds a pair of boundary fixed points $E_0$ and $E_1$  located at $(0,0)$ and $(1,0)$, respectively. Another possible boundary fixed point $E_2$ on the $y$-axis at $(0,y_2)$ gets eliminated as the existence of this requires $d = 0$.

For $E_0(0,0)$, the Jacobian turns out to be 
\begin{gather}
J(E_0) = \begin{bmatrix}
0 & 0\\ 0 & -d
\end{bmatrix}
\end{gather}
Therefore, near the origin, the vector field is vertical and attractive, while even the slightest perturbation will cause the system to move along the x-axis.

 $E_1(1,0)$ is the next boundary fixed point to consider whose Jacobian turns out to be 
 \begin{gather}
 J(E_1) = \begin{bmatrix}
 -(1-e^{-a}) & -1\\ 0 & \epsilon - d
 \end{bmatrix}
 \end{gather}
 This implies that trace $T(E_1) = \epsilon - d - 1 + e^{-a}$ and determinant $D(E_1) = (d-\epsilon)(1- e^{-a})$. For $d > \epsilon$, $T(E_1) < 0$ and $D(E_1) > 0$ which makes $E_1(1,0)$ a stable fixed point. This means for $d> \epsilon$ condition, any initial set of population in the basin of attraction of $E_1$ will lead to predator extinction over time. 
However, for $d <\epsilon$ and $D(E_1) < 0$, $E_1$ becomes a saddle type unstable fixed point and more on this will follow in later subsections. Another key point to ponder upon is all these stability wise results till now are invariant of choice for Allee parameter $a$. However, there may exist a choice $a = \tilde{a} = -\ln(1 + d - \epsilon)$, which makes $T^2(E_1) = 4D(E_1)$, and by choosing $a$ in either side of $\tilde{a}$ will transform $E_1(1,0)$ from stable node to stable spirals and may effect extinction time a little but the equilibrium nature will not change as long as $d > \epsilon$. 
 
\subsubsection{Interior Fixed Point}
The interior fixed points will be addressed by $E^*(x^*,y^*)$ which are solutions of the intersection of null-clines given by 
\begin{subequations}
	\begin{align}
	& x^* = \frac{d}{\epsilon}\\
	& y^* = \bigg(1 - \frac{d}{\epsilon}\bigg)\bigg(1 - e^{-\frac{ad}{\epsilon}}\bigg)
	\end{align}
\end{subequations} 
It is noted that $E^*(x^*,y^*)$ exists in positive quadrant only for $d < \epsilon$ which in turn imposes $ x^*<1$. The Jacobian at interior fixed point turns out to be
\begin{gather}
J(E^*(x^*,y^*)) = \begin{bmatrix}
x^*\big[e^{-ax^*}(1+a-ax^*)-1\big] & -x^*\\ \epsilon y^* & 0\end{bmatrix}
\end{gather}
 One obtains the trace $T(E^*)$ and determinant $D(E^*)$ as
 \begin{gather}
 \begin{split}
 & T(E^*) = \frac{d}{\epsilon}\big[e^{-\frac{ad}{\epsilon}}(1+a-\frac{ad}{\epsilon})-1\big]\\
 & D(E^*) = \epsilon x^*y^* 
 \end{split}
 \end{gather}
It may be observed that $ D(E^*) = \epsilon x^*y^* > 0$, which implies that $E^*(x^*,y^*)$ is never a saddle type fixed point. Its stability depends solely upon the associated value of $T(E^*)$. 
The equilibrium point  $E^*(x^*,y^*)$  is locally asymptotically stable if the trace $T(E^*)<0$, and it is unstable if the trace  $T(E^*)>0$. The value of the parameter $a$ at which $T(E^*)$ vanishes, denoted by $a_0$, is a solution of a transcendental equation:
\begin{gather}
a_0\bigg(1-\frac{d}{\epsilon}\bigg) = e^\frac{a_0 d}{\epsilon} -1 
\end{gather}
While the above equation is always satisfied by $a_0 = 0$, this trivial solution corresponds to the case where the Allee effect is not present.
However, a non-trivial positive root $a_0 = a_H$ may exist at which the trace vanishes and the determinant is also positive, enabling an onset of the Hopf bifurcation. Further analytical results may be drawn in terms of the Hopf bifurcation by asking when the non-trivial $a_H$ gets separated from the trivial one at $a_0 = 0$. Existence of $a_H$ is guaranteed by the condition
\begin{gather}
\frac{d}{d a_0}\bigg(e^{a_0 d/\epsilon}\bigg)_{a_0 = 0}  < \frac{d}{d a_0}\bigg(\big(1-\frac{d}{\epsilon})a_0 + 1\bigg)_{a_0 = 0}
\end{gather}
which gets simplified to
\begin{gather}
\frac{d}{\epsilon} < \frac{1}{2}
\end{gather}

So, it may be concluded that for $\frac{d}{\epsilon} > 1$, $E_1(1,0)$ is the stable fixed point leading to the extinction of predators while for $\frac{1}{2} < \frac{d}{\epsilon} <1$, $E^*(x^*,y^*)$ is the stable fixed point and finally for $ 0 < \frac{d}{\epsilon} < \frac{1}{2} $ range, there exists a value $a = a_H$ where stability of $E^*(x^*,y^*)$ goes away vide the Hopf bifurcation, giving rise to a stable limit cycle in the process. 
 
\subsection{Numerical Analysis}
To complete the discussion on the dynamical aspects of the model under consideration, the phase trajectories for four different initial conditions with varying sets of parameters of  $d$ and $a$ have been drawn for the fixed value of $\epsilon = 1$. Firstly, by setting $d= 1.5 > \epsilon$, the phase portrait in Figure 1(a) shows the predator population extinct over time. Whereas, for $d = 0.75$, for which $\frac{1}{2} < \frac{d}{\epsilon} < 1$, the phase trajectories converge to the corresponding interior fixed point $E^*(x^*,y^*)$ as shown in Figure 1(b) which ensures the coexistence of predator and prey populations. The population dynamics topology for parameter value $\frac{d}{\epsilon} > \frac{1}{2}$ remains unaffected by the selection of the Allee parameter $a$. Finally, to investigate the domain $\frac{d}{\epsilon} < \frac{1}{2}$, setting the value of $d = 0.4$, the numerical value of $a = a_H$ is found to be 1.9067214. Thus, the phase portraits for $a = 2.0$ and $a = 1.8$ have been observed in Figures 1(c) and 1(d), respectively. In Figure 1(c), all trajectories spiral into the interior fixed point. In contrast, in Figure 1(d), all trajectories converge onto the limit cycle born out of the supercritical Hopf bifurcation, ensuring oscillatory behaviour of both predator as well as prey populations. It is also to be noted that the existence of the limit cycle is dependent on the corresponding Allee parameter value $a$, in turn highlighting the importance of the Allee effect on population dynamics.
\begin{figure}[H]
	\begin{center}
		\includegraphics[height=15cm, angle=0]{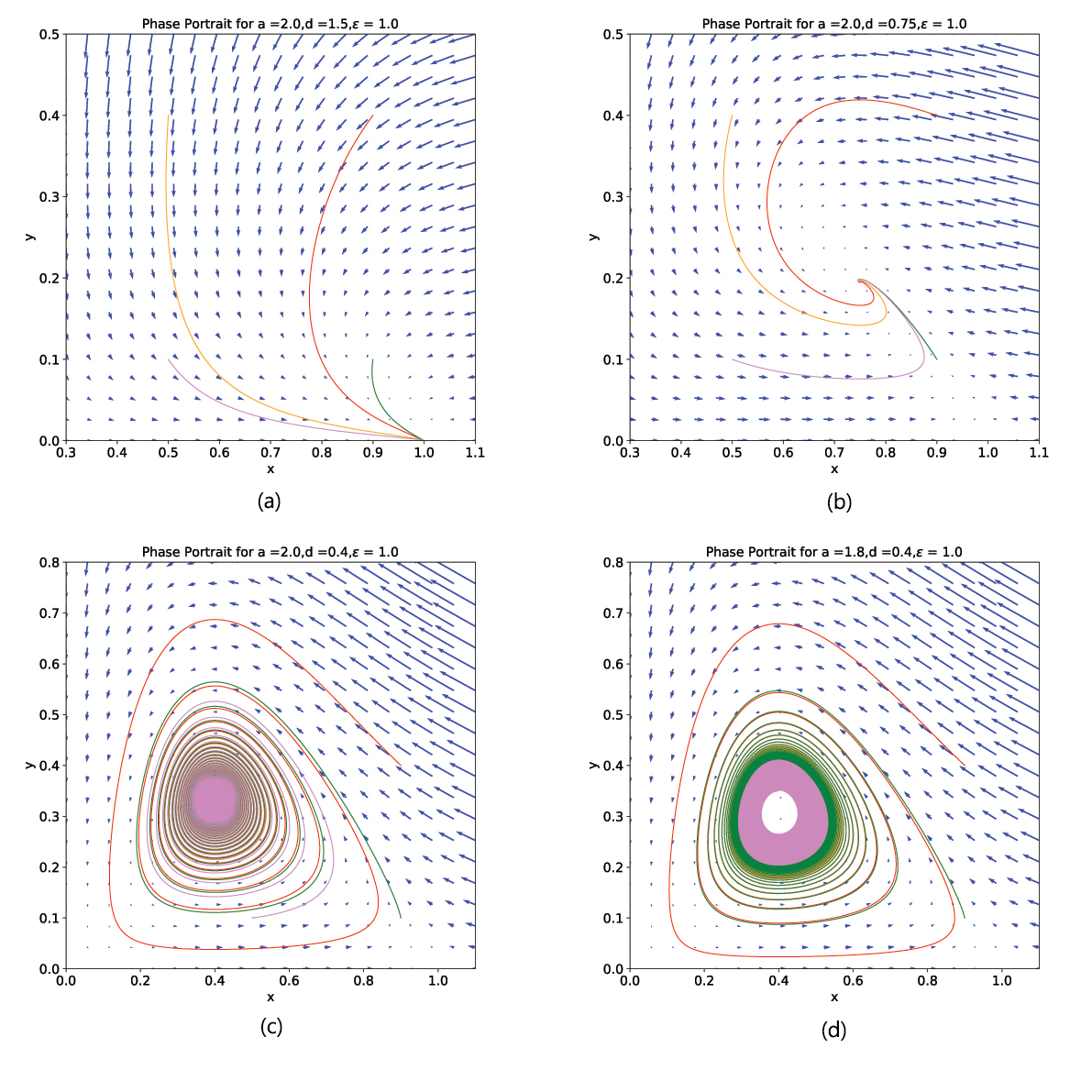}
		\caption{Phase portrait containing four phase trajectories for each parameter set has been obtained. Each set has $\epsilon = 1.0$. In Figure 1(a), choice of parameter $d = 1.5, a = 2.0$ shows prey population extinct over time.Phase portrait in Figure 1(b) has parameter value $d = 0.75, a = 2.0$ showing stable interior fixed point. Figures 1(c) and 1(d) have been drawn for parameter values $d = 0.4, a = 2.0$ and $d = 0.4, a= 1.8$ reflecting the occurrence of Hopf-bifurcation for $1.8 < a < 2.0$.}
		\label{Figure 1}
	\end{center}
\end{figure}

\section{The Allee Effect in presence of Holling type-II functional response}
Having explored the dynamical implications of the Allee effect on prey populations under the Holling type-I functional response, analogous effects under the Holling type-II functional response will be examined in the following section.

The model under study is given by
\begin{subequations}
	\begin{align}
	\frac{dN}{dT} &= \bar{r} N\bigg(1 - e^{-\bar{a} N}\bigg)\bigg(1 - \frac{N}{K}\bigg) - \frac{ANP}{1+\bar{h}N}\\
	\frac{dP}{dT} &= \bar{\epsilon}\frac{ANP}{1+\bar{h}N} - DP
	\end{align}
\end{subequations}
where, $\bar{r}, A, \bar{h}, \bar{\epsilon}, D$ are all positive definite constants. Again,this model will also be required to be non-dimensionalized by rescaling variables as:
\begin{gather}
x = \mu N, y = \nu P, t = \beta T
\end{gather}
After few simplifications, this substitution leads to 
\begin{subequations}
	\begin{align}
	\frac{dx}{dt} &=  \frac{\bar{r}}{\beta} x\bigg(1 - e^{-\frac{\bar{a} x}{\mu}} \bigg)\bigg(1 - \frac{x}{\mu K}\bigg) - \frac{A}{\beta \nu} \frac{xy}{1+\frac{A\bar{h}}{\mu}x}\\
	\frac{dy}{dt} &= \bar{\epsilon}\frac{A}{\beta \mu}\,\frac{xy}{1+\frac{A\bar{h}}{\mu}x} - \frac{D}{\beta}\,y 
	\end{align}
\end{subequations}
The normalization parameters then chosen to be 
\begin{gather*}
\mu = \frac{1}{K}, \beta = \bar{r}, \nu = \frac{A}{\bar{r}}
\end{gather*}
which leads to the non-dimensional model:
\begin{subequations}
	\begin{align}
	\frac{dx}{dt} &= x\big(1 - e^{-ax}\big)\big(1 - x\big) - \frac{xy}{1+hx}\\
	\frac{dy}{dt} &= \epsilon \frac{xy}{1+hx} - d y
	\end{align}
\end{subequations}
where, few parameters have been re-labeled as $ a = \frac{\bar{a}}{\mu},\epsilon = \frac{\bar{\epsilon} AK}{\bar{r}}, d = \frac{D}{\bar{r}}$, and $h=A\bar{h}K$. Thus, the Allee function in the non-dimensional form turns out to be $\alpha(x) = 1 - e^{-ax}$ same as before.

\subsection{Local stability Analysis of the Model}
Following the last section, firstly, all available fixed points will be identified in the next subsection, and stability around them will be discussed.
\subsubsection{ Boundary Fixed points}
The system exhibits two boundary equilibrium points, $E_0 = (0,0)$ and $E_1 = (1,0)$. Furthermore, the fixed point $E_2$, which would normally be situated on the y-axis at $(0, y_2)$, does not exist due to the physical constraint that $d\neq0$.

The Jacobian for two boundary ﬁxed points is similar to the last section's.
\begin{gather}
J(E_0) = \begin{bmatrix}
0 & 0\\ 0 & -d
\end{bmatrix}
\end{gather}
\begin{gather}
J(E_1) = \begin{bmatrix}
-(1-e^{-a}) & -\frac{1}{1+h}\\ 0 & \frac{\epsilon}{1+h} - d
\end{bmatrix}
\end{gather}
This shows that $J(E_0)$ has the same form as the last one while the determinant and trace at $E_1(1,0)$ are given by
\begin{gather}
\begin{split}
D(E_1) & = (1-e^{-a})(d-\frac{\epsilon}{1+h}) \\
T(E_1) & = -(1-e^{-a})+ \frac{\epsilon}{1+h} - d
\end{split}
\end{gather}
For $\epsilon < d(1+h) = \epsilon_0$, the fixed point $E_1(1, 0)$ exhibits stability, confirmed by the fact that the determinant $D(E_1)$ is positive and the trace $T(E_1)$ is negative, causing the predator population to eventually become extinct.
However, as will be observed in the next subsection, $\epsilon >  \epsilon_0$ condition would lead to some interesting dynamical aspects.

\subsubsection{Interior Fixed Point}
The interior fixed points will be addressed by $E^*(x^*,y^*)$ which are the solutions of the intersection of null-clines given by 
\begin{subequations}
	\begin{align}
	& x^* = \frac{d}{\epsilon-dh}\\
	& y^* = (1 - e^{-ax^*})(1-x^*)(1+hx^*) 
	\end{align}
\end{subequations}
The positivity of both $x^*$ and $y^*$ once again constraints $0< x^*<1$ which in turn imposes, $\epsilon >  \epsilon_0$. Thus, the birth of the interior fixed point $E^*(x^*,y^*)$ would destabilize the boundary fixed point $E_1(1,0)$. Now, the Jacobian at the interior fixed point $E^*(x^*,y^*)$ turns out to be
\begin{gather}
J(E^*(x^*,y^*)) = \begin{bmatrix}
x^*\big[\frac{hy^*}{(1+hx^*)^2} +e^{-ax^*} (1+a-ax^*) -1\big] & -\frac{x^*}{1+hx^*}\\ \frac{\epsilon y^*}{(1+hx^*)^2} & 0\end{bmatrix}
\end{gather}
which gives trace $T(E^*)$ and determinant $D(E^*)$ as
\begin{gather}
\begin{split}
& T(E^*) = x^*\big[\frac{hy^*}{(1+hx^*)^2} +e^{-ax^*} (1+a-ax^*) -1\big] \\
& D(E^*) = \frac{\epsilon x^*y^*}{(1+hx^*)^3}
\end{split}
\end{gather}
The positive determinant 
$ D(E^*) = \frac{\epsilon x^*y^*}{(1+hx^*)^3}>0 $ indicates that $E^*(x^*,y^*)$ is never a saddle point.
Its stability depends solely upon the associated value of $T(E^*)$. To study the Hopf bifurcation in this regard, the associated parameter value of $a$ may be denoted as $a_0$ which satisfies:  
\begin{gather}
1-e^{-a_0 x^*} =\frac{hy^*}{(1+hx^*)^2} + a_0 e^{-a_0 x^*} (1-x^*)
\end{gather}
By replacing $y^* = (1 - e^{-a_0x^*})(1-x^*)(1+hx^*)$ in last equation, one obtains
\begin{gather}
e^{a_0 x^*} -1 =\frac{(1-x^*)(1+hx^*)}{1+2hx^*-h} a_0
\label{tran1}
\end{gather}
where, it must be noted that $x^* = d/(\epsilon-dh)$. This leads to a transcendental equation for $a_0$, which has a trivial solution $a_0=0$ regardless of the values of the other parameters. 
However, to obtain a non-trivial solution $a_0 = a_H$, one must have
\begin{gather}
\begin{split}
\frac{d}{da_0}\bigg(e^{a_0x^*} -1\bigg)_{a_0 = 0} = x^* \le \frac{(1-x^*)(1+hx^*)}{1+2hx^*-h} 
\end{split}
\end{gather}
The above equation may be solved for $x^*$ and then substituting $x^* = d/(\epsilon-dh)$, one can find the critical value for $\epsilon$ above with the non-trivial $a_0 = a_H$ root exists:
\begin{equation}
\label{critical_eps}
\epsilon \ge \epsilon_{\rm critical} = dh\bigg(1+\frac{3}{h-1+\sqrt{h^2+h+1}}\bigg)
\end{equation}
To summarize, the population dynamics evolve for different domains of parameter value $\epsilon$. For $\epsilon < \epsilon_0$, the predator population gets extinct over time, while beyond $\epsilon > \epsilon_0$, the population coexistence occurs. For the later mentioned domain, there exists a value $\epsilon_{\rm critical}$, such that choosing $\epsilon_0 < \epsilon < \epsilon_{\rm critical}$ populations stabilize to an equilibrium value, but for $\epsilon > \epsilon_{\rm critical}$, a new possibility appears, and it is possible to determine a value $a = a_H$, which ensures an oscillatory population pattern for any choice of $a < a_H$. It is worth reminding that the strength of the Allee effect decreases with the increment of the Allee parameter $a$, and the oscillatory behaviour is imposed only in the presence of the Allee parameter, which makes it ecologically important.

\subsection{Numerical Analysis}
For the numerical study, the parameter values of $h$ and $d$ are set equal to 1.6 and 0.6, respectively, which gives corresponding $\epsilon_0 = 1.56$ and $\epsilon_{\rm critical} = 1.9629380029920656$. Following the strategy depicted in the previous subsection, the numerical analysis of the current model will be subdivided into three parts: $\epsilon < \epsilon_0$, $\epsilon_0 < \epsilon < \epsilon_{\rm critical}$ and then $\epsilon > \epsilon_{\rm critical}$.
\begin{figure}[H]
	\begin{center}
	\label{Figure 2}
		\includegraphics[height=7cm, angle=0]{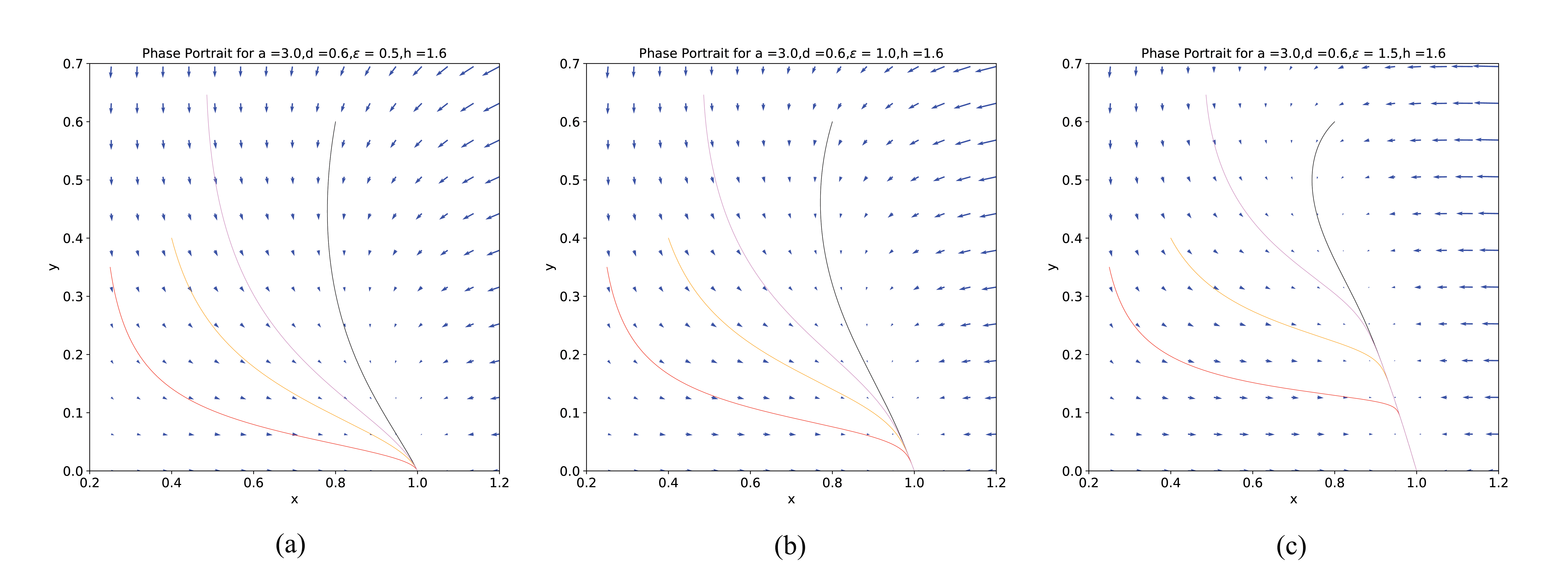}
		\caption{Phase portraits drawn in $\epsilon<\epsilon_0 = 1.56$ parameter domain. In each case other parameters were kept fixed at $a = 3.0$, $d = 0.6$ and $h= 1.6$ while choosing (a) $\epsilon = 0.5$ (b) $\epsilon = 1.0$ and (c) $\epsilon = 1.5$, for respective figures.}
		\label{Figure 2}
	\end{center}
\end{figure}

For the $\epsilon <\epsilon_0$ domain, the values of $\epsilon$ are selected as $0.5, 1.0, \text{ and } 1.5$. The corresponding phase portraits for the same are drawn in Figures \ref{Figure 2}(a), (b), and (c), respectively. In all these cases, the Allee parameter $a$ is set to $3.0$. 
As discussed in the previous subsection, the predator population becomes extinct over time in each scenario. 

\begin{figure}[H]
	\begin{center}
	\label{Figure 3}
		\includegraphics[height=7cm, angle=0]{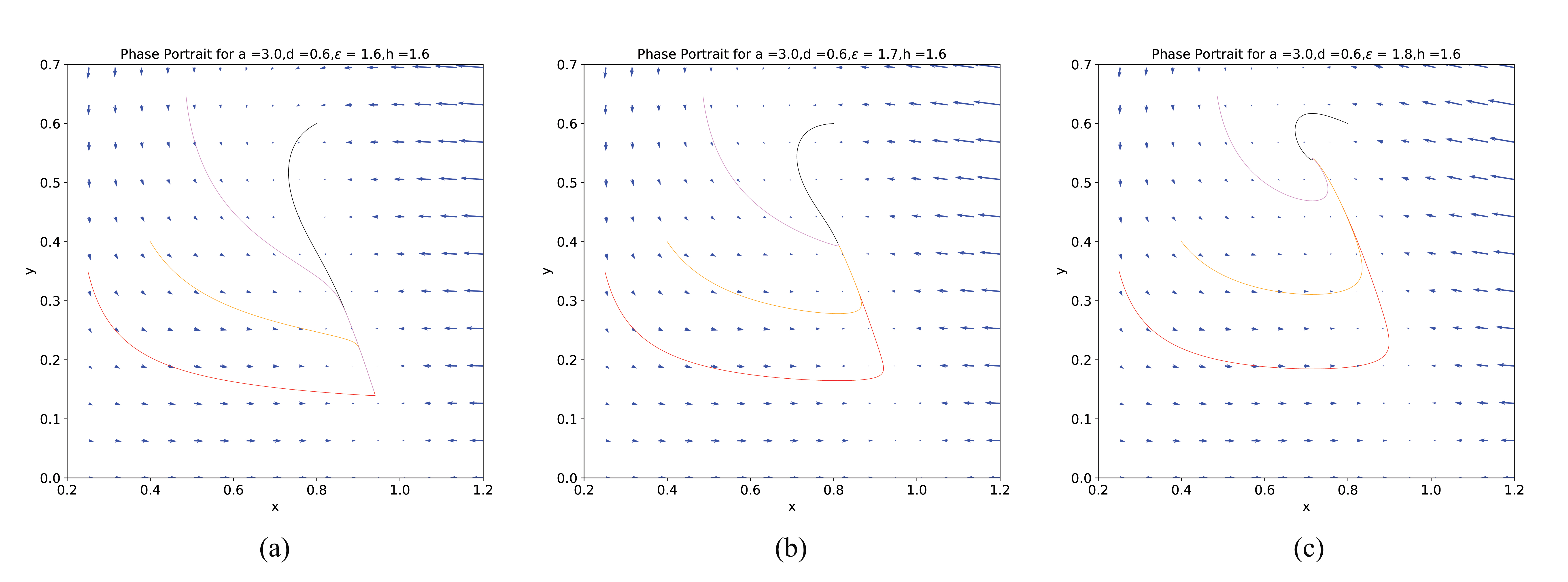}
		\caption{Phase portraits drawn in $\epsilon_0 (= 1.56) < \epsilon< \epsilon_{\rm critial} (\approx 1.963)$ parameter domain. In each case other parameters were kept fixed at $a = 3.0$, $d = 0.6$ and $h= 1.6$ while choosing (a) $\epsilon = 1.6$ (b) $\epsilon = 1.7$ and (c) $\epsilon = 1.8$, for respective figures.}
		\label{Figure 3}
	\end{center}
\end{figure}

For the $\epsilon_0 <\epsilon <\epsilon_{\rm critical}$ domain, $\epsilon$ is chosen as $1.6,1.7$ and $1.8$, while the Allee parameter is fixed at $a=3.0$. Three phase portraits with different initial conditions are depicted in Figures \ref{Figure 3}(a), (b), (c), respectively. 
The convergence of trajectories to the interior fixed point $E^*(x^*,y^*)$ demonstrates that the two species coexist within this domain.
\begin{figure}[H]
	\begin{center}
		\label{Figure 4}
		\includegraphics[height=5.5cm, angle=0]{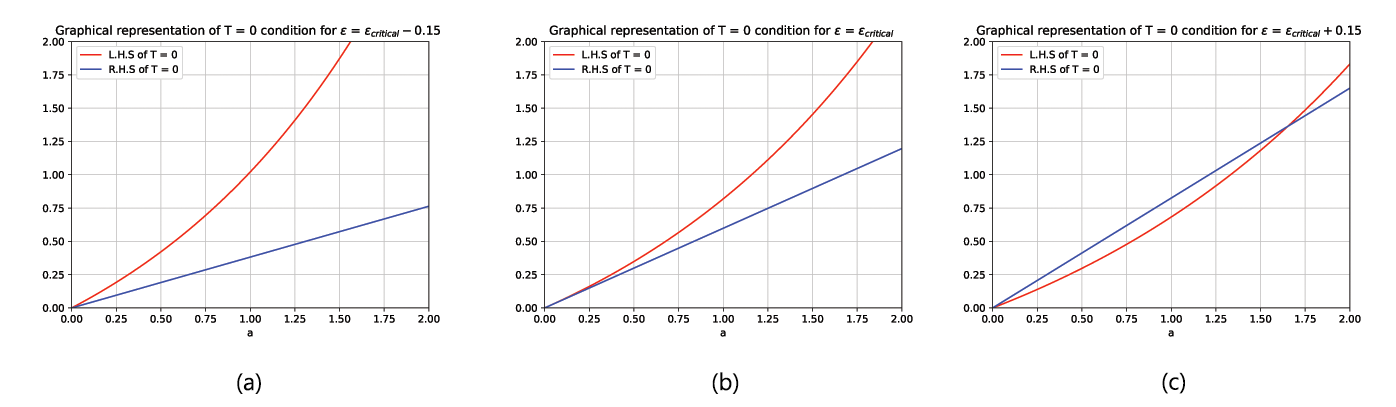}
		\caption{Graphical solution for Trace $=0$ condition with  $h = 1.6, d=0.6$ and $\epsilon = \epsilon_{\rm critical} - 0.15, \epsilon_{\rm critical}, \epsilon_{\rm critical} +0.15$, respectively}
		\label{Figure 4}
	\end{center}
\end{figure}

To investigate the possibility of a Hopf bifurcation in the $\epsilon > \epsilon_{\rm critical}$ domain, a numerical case study was conducted using the same parameter values: h = 1.6 and d = 0.6.
For a graphical representation of $a_H$, the left and the right-hand sides of equation (\ref{tran1}) as a function of $a$ have been plotted in Figure (\ref{Figure 4}) with $\epsilon$ having three values: $\epsilon_{\rm critical} -0.15$,  $\epsilon_{\rm critical}$ and  $\epsilon_{\rm critical} + 0.15$, respectively. As it can be observed in Figure \ref{Figure 4}(c), choosing $\epsilon > \epsilon_{\rm critical}$, it is possible to find a nontrivial root of $a = a_H =  2.58882981$, 
suggesting the possibility of a Hopf bifurcation. 
Therefore, when $\epsilon > \epsilon_{\rm critical}$ and $a < a_H$, the system is expected to exhibit oscillatory behavior in the predator-prey populations.

\begin{figure}[H]
	\begin{center}
	\label{Figure 5}
		\includegraphics[height=6.3cm, angle=0]{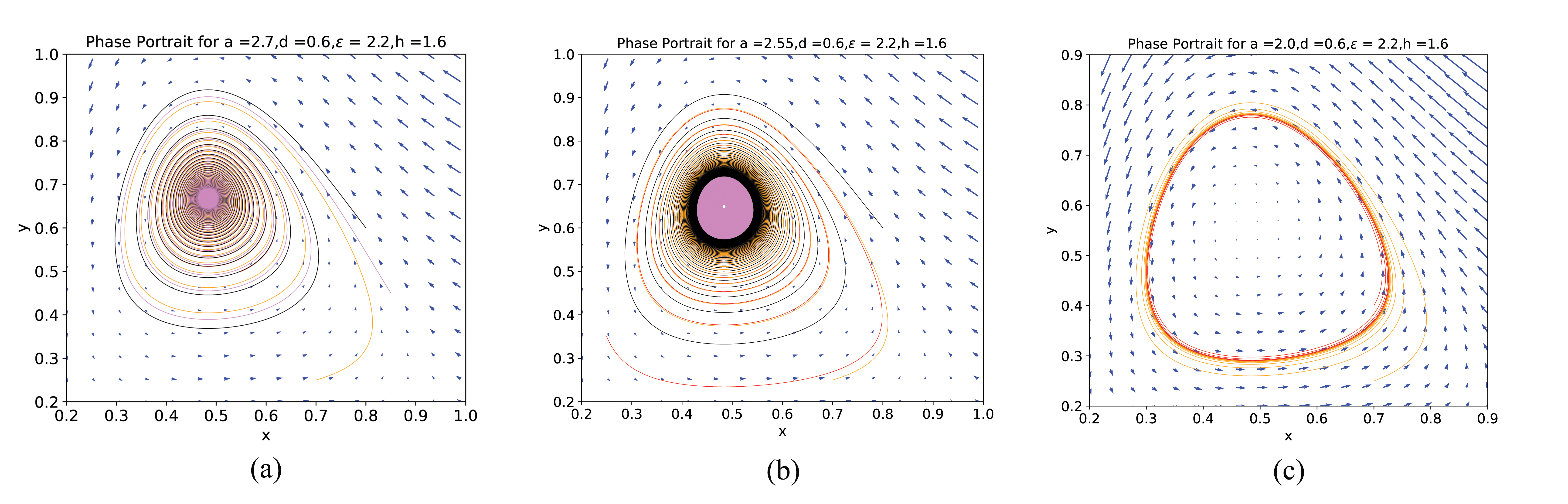}
		\caption{Phase portrait for different parameter values: (a) and (b) for $\epsilon > \epsilon_{\rm critical}$ with $a>a_H$ and $a<a_H$,  respectively showing the emergence of stable limit cycle via Hopf bifurcation, (c) represents a relatively larger stable limit cycle away from Hopf condition}
		\label{Figure 5}
	\end{center}
\end{figure}

Finally, the phase portraits for $\epsilon > \epsilon_{\rm critical}$ have been drawn for different Allee parameter values, as shown in Figure \ref{Figure 5} for $a = 2.7, 2.55$, and $2.0$, respectively. \ref{Figure 5}(a) and (b) show phase trajectories just before and after the birth of a stable limit cycle via supercritical Hopf bifurcation.  For a more enlarged limit cycle, the phase portrait has been dawned in Figure \ref{Figure 5}(c) away from $a = a_H$.
These phase portraits reveal the emergence of a stable periodic orbit around the unstable interior equilibrium point for $a<a_{H}$, which suggests a mechanism for the long-time oscillatory behavior of the species.

\section{Conclusion}
The role of the Allee effect on prey in population dynamics typically destabilizes predator-prey dynamics, and is widely studied in both empirical and theoretical contexts. The Gause-type predator-prey model, incorporating logistic prey growth with predator functional response, has been widely studied~\cite{Courchamp2000, Kent2003, Zhou2005, Boukal2007}. Due to the Allee effect on the prey population, both the predator and prey populations can coexist at a stable state or an oscillatory state. In this manuscript, a novel function representing the Allee effect is proposed and validated against the ecological criteria to ensure its biological relevance and applicability. The impact of the Allee effect on prey is analyzed in predator-prey models with Holling type-I and type-II functional responses. Initially, an appropriate scaling simplifies both systems, resulting in topologically equivalent predator-prey models with fewer parameters. The stability of the various equilibria in these systems is also discussed.

For a Holling type-I functional response, when the predator's death rate exceeds its food conversion rate ($d>\epsilon$), the predator-free equilibrium is asymptotically stable, regardless of the Allee effect. When $d<\epsilon$, a single interior equilibrium point exists. 
If $\frac{1}{2}<\frac{d}{\epsilon}<1$, all phase trajectories converge to this point, ensuring the long-term coexistence of predator and prey populations. However, when $\frac{d}{\epsilon}<\frac{1}{2}$, the interior equilibrium is stable if the Allee parameter, $a$, exceeds a critical threshold, $a_H$. A Hopf bifurcation occurs at $a = a_H$, destabilizing the equilibrium and leading to the emergence of a stable periodic orbit that encompasses the unstable equilibrium point. 

For the model with the Holling type-II functional response, the analytical results indicate that the boundary equilibrium $E_1(1,0)$ is asymptotically stable as long as 
$\epsilon<\epsilon_0$ is satisfied. 
When $\epsilon>\epsilon_0$, $E_1$ loses its stability, the system transits to a state characterized by an interior fixed point, which is never a saddle point.
In this scenario, the stable coexistence of prey 
and predator occurs at the interior fixed point 
when $\epsilon$ is less than the critical value,
$\epsilon_{\rm critial}$.
A Hopf bifurcation occurs when 
$\epsilon>\epsilon_{\rm critial}$,
destabilizing the steady state of coexistence.
This bifurcation leads to a new dynamic regime characterized by the stable oscillatory coexistence of the predator and prey populations.
In such a predator-prey system, the strength of the Allee effect diminishes as the Allee parameter increases, but its presence is essential for the emergence of the oscillatory behavior.

\vskip 1cm
\noindent {\bf Acknowledgements}\\\\ 
Tanmay Das likes to thank Sananda Basak for continuous support and encouragement.



\begin{thebibliography}{99}
\bibitem{May1973}May, R.M. (1973) Stability and Complexity in Model Ecosystems.
Princeton University Press, Princeton, NJ.
\bibitem{Crawley1992}Crawley, M.J. (Ed.) (1992)
Natural Enemies: The Population Biology
of Predators, parasites and Diseases. Blackwell Scientiﬁc, London.
\bibitem{Mueller2000}Mueller, L.D., Joshi, A. (2000) Stability in Model Populations.
Princeton University Press, Princeton, NJ.
\bibitem{Neubert2002}Neubert, M.G., Klepac, P., Driessche, P. (2002) Stabilizing dispersal
delays in predator–prey metapopulation models. Theor. Popul. Biol. 61, 339–347.
\bibitem{Berryman1992}Berryman A. A. (1992) The  Origins  and  Evolution  of  Predator-Prey  Theory. Ecology 73, 1530–1535.
\bibitem{Kuang1998} Kuang  Y.,  Beretta  E. (1998) Global   qualitative  analysis  of  a  ratio-dependent  predator–prey  system. Journal of Mathematical Biology 36, 389–406.
\bibitem{strogatz}  Strogatz S. H. (1994) Nonlinear dynamics and Chaos, with applications to Physics, Biology, Chemistry and Engineering, Advanced Book Program, Perseus Books.
\bibitem{selkov} Sel'kov E. E. (1968) Self-oscillationsin Glycolysis, European J. Biochem. 4, 79.
\bibitem{Das} Das T., Acharyya M. (2021) Transient behavior towards the stable limit cycle in the Sel'kov model of Glycolysis: A physiological disorder, Physica A, 567, 125684.
\bibitem{Pita} E. Diz-Pita, M. V. Otero-Espinar. (2021) Predator–Prey Models: A Review of Some Recent Advances, Mathematics  9(15), 1783

\bibitem{Lotka} Lotka, A.J.  (1926)  Elements of Physical Biology. Elem. Phys. Biol. 82, 341-343
\bibitem{Volterra} Volterra, V. (1926) Variations and ﬂuctuations of the number of individuals in animal species living together. J. Cons. Perm. Int. Ent. Mer. 3, 3–51.

\bibitem{Allee} Allee, W.C. (1931) Animal Aggregations, a Study in General Sociology. University of Chicago Press, Chicago, USA.


\bibitem{Dennis1989} Dennis B. (1989) Allee effects: population growth, critical density and the chance of extinction. Natural Resource Modeling, 3, 481-538.
\bibitem{Courchamp2008} Courchamp F., Berec L., Gascoigne J. (2008) Allee effect in ecology and 
conservation, Oxford University Press, Oxford.
\bibitem{Gascoigne2004} Gascoigne J. and Lipcius R. (2004) Allee effects driven by predation. Journal of Applied Ecology, 41, 801-10.
\bibitem{Courchamp2006}Courchamp F., Angulo E., Rivalan P., et al. (2006) Rarity value and species extinction: the anthropogenic Allee effect. Plos Biology, 4, 2405-10.
\bibitem{Stephens1999}Stephens P.A. and Sutherland WJ. (1999) Consequences of the Allee effect for behaviour, ecology and conservation. Trends in Ecology \& Evolution, 14, 401-5.
\bibitem{Sinclair1998}Sinclair, A.R.E., Pech, R.P., Dickman, C.R., Hik, D., Mahon, P., Newsome, A.E.. ``Predicting effects of predation on conservation of endangered prey". Conserv. Biol. 12, 564–575, , 1998.
\bibitem{Courchamp1999} Courchamp, F., Clutton-Brock, T., Grenfell, B. (1999) Inverse density dependence and the Allee effect. Trends Ecol. Evol. 14, 405–410.
\bibitem{Courchamp2000}Courchamp P., Grenfell B.T., and Clutton-Brock T.H. (2000) Impact of natural enemies on obligately cooperative breeders. Oikos, 91, 311-22.
\bibitem{Kent2003} Kent A., Doncaster C.P., Sluckin T.  (2003) Consequences for predators of rescue and Allee effects on prey. Ecological Modelling 162, 233-45, 2003.
\bibitem{Zhou2005}Zhou S.R., Liu YE, and Wang G.  (2005) The stability of predator-prey systems subject to the Allee effects. Theoretical Population Biology, 67, 23-31.

\bibitem{Wang2010} Wang J., Shi J., Wei J. (2010)
Predator-prey system with strong Allee effect in prey. J. Math. Biol., 62, 291–331.
\bibitem{Boukal2007} Boukal D.S., Sabelis M.W., Berec L. (2007) How predator functional responses and Allee effects in prey affect the paradox of enrichment and population collapses. Theoretical
Population Biology, 72, 136–147.

\bibitem{Shi2006} Shi J., Shivaji R. (2006) Persistence in reaction diffusion models with weak Allee effect. J. Math. Biol., 52, 807-829.
\bibitem{Wang2001} Wang M., Kot M. (2001) Speeds of invasion in a model with strong or weak Allee effect. Math. Biosci. 171, 83-97.



\bibitem{Celik2008} Celik C., Merdan H., Duman O., Akin O. (2008) Allee effects on population dynamics with delay. Chaos Solitons Fractals, 37,
65–74.
\bibitem{Celik2009} Celik C., Duman O. (2009) Allee effect in a discrete-time predator–prey system. Chaos Solitons Fractals 2009, 90, 1952–1956.
\bibitem{Merdan2009} Merdan H., Duman O., Akın Ö., Çelik C. (2009) Allee effects on population dynamics in continuous (overlaping) case. Chaos Solitons Fractals, 39, 1994–2001.
\bibitem{Gonzalez2010} Gonzalez-Olivares E., Mena-Lorca J., Rojas-Palma A. (2010) Dynamical complexities in the Leslie-Gower predator-prey model as consequences of the Allee effect on prey. Appl. Math. Model. 35, 366–381.
\bibitem{Gonzalez2019} Gonzalez-Olivares E., Cabrera-Villegas J., Cordova-Lepe F., Rojas-Palma A. (2019) Competition among Predators and Allee Effect on Prey, Their Influence on a Gause-Type Predation Model. Math. Probl. Eng. 2019, 3967408.
\bibitem{Flores2014} Flores D. J., Gonzalez-Olivares E. (2014) Dynamics of a predator-prey model with Allee effect on prey and ratio-dependent functional response. Ecological Complexity, 18, 59-66. 
\bibitem{Sasmal2018} Sasmal S.K., (2018) Population dynamics with multiple Allee effect induced by fear factors-A mathematical study on prey-predator interactions. Appl. Math. Model. 64, 1–14.
\bibitem{Sen2019} Sen D., Ghorai S., Banerjee M. (2019) Allee Effect in Prey versus Hunting Cooperation on Predator - Enhancement of Stable Coexistence. International Journal of Bifurcation and Chaos 29, 06, 1950081.
\bibitem{Sen2021} Sen D., Ghorai S., Sharma S., Banerjee M. (2021) Allee effect in prey's growth reduces the dynamical complexity in prey-predator model with generalist predator. Appl. Math. Mode. 91, 768-790.
\bibitem{Ghosh2017} Ghosh K., Biswas S., Samanta S., Tiwari P.K., Alshomrani A.S., and Chattopadhyay J. (2017) Effect of Multiple Delays in an Eco-Epidemiological Model with Strong Allee Effect. International Journal of Bifurcation and Chaos 27, 11, 1750167
\bibitem{Kumar2022} Kumar U., Mandal P. S. (2022) Role of Allee effect on prey-predator model with component Allee effect for predator reproduction. Math. \& Comp. Simu. 193, 623-665.
\bibitem{Han2003} Han R., Dey S., Banerjee M. (2023) Spatio-temporal pattern selection in a prey–predator model with hunting cooperation and Allee effect in prey. Chaos Solitons Fractals, 171, 113441. 

\bibitem{Merdan} Merdan, H. (2010) Stability analysis of a Lotka-Volterra type predator-prey system involving Allee effects. Anziam J.  52, 139–145.
\bibitem{Berec2007} Berec L., Angulo E., Courchamp F. (2007) Multiple Allee effects and population management. Trends in Ecology \& Evolution, 22, 4, 185-191. 

\bibitem{Holling1959} Holling C.S. (1959) Some characteristics of simple types of predation and parasitism. Canadian Entomologist, 91, 385-98.

\bibitem{Holling1965} Holling C.S. (1965) The functional response of predators to prey density and its role in mimicry and population regulation. Mem. Entomol. Soc. Can., 97(S45), 5–60


\bibitem{Seo2008}Seo G.,  Kot M. (2008) A comparison of two predator-prey models with Holling’s type I functional response, Math. Biosci. 212, 161–179.

\bibitem{Rosenzweig1963} Rosenzweig, M.L. and MacArthur, R. (1963) Graphical Representation and Stability Conditions of Predator-Prey Interactions. The American Naturalist, 97, 209-223.
\bibitem{Zhang2006} 	Zhang, C.M., Chen, W.C., Yang, Y. (2006) Periodic Solutions and Global Asymptotic Stability of a Delayed Discrete Predator-Prey System with Holling II Type Functional Response. Journal of Systems Science and Complexity, 19, 449-460.
\bibitem{Zu2010} Zu J., Mimura M. (2010) The impact of Allee effect on a predator-prey system with Holling type II functional response. Appl. Math. Comp. 217, 3542-3556.
\bibitem{Zu2013} Zu J. (2013) Global qualitative analysis of a predator-prey system with Allee effect on the prey species. Math. \& Comp. Simu. 94, 33-54.
\bibitem{McLellan2010} McLellan B.N., Serrouya R., Wittmer H.U., Boutin S. (2010) Predator-mediated Allee effects in multi-prey systems. Ecology 1, 286-292. 
\bibitem{Halder2019} Halder S., Bhattacharyya J., Pal S. (2019)  Comparative studies on a predator-prey model subjected to fear and Allee effect with type I and type II foraging.  Jour. Appl. Math. \& Comp. 62, 93-118. 





%


\end{thebibliography}
\end{document}